\documentclass[a4paper,fleqn]{cas-dc}

\usepackage{dcolumn}
\usepackage{bm}
\usepackage{float}
\usepackage{xcolor}
\DeclareMathOperator{\sgn}{sgn}
\usepackage{flafter}
\def\fnum@figure{\figurename\thefigure}
\renewcommand{\figurename}{Fig.}

\usepackage[numbers]{natbib}

\begin{document}
\let\WriteBookmarks\relax
\def\floatpagepagefraction{1}
\def\textpagefraction{.001}
\shorttitle{Broken spatial symmetry and non-reciprocity}
\shortauthors{A. Maiti et~al.}

\title[mode = title]{Probing the Broken Spatial Symmetry of a Stratified Medium with Structured Light}

\author[1]{Arani Maiti}
\date{\today}
\author[1]{Sauvik Roy}
\author[1]{Nirmalya Ghosh}
\author[1]{Ayan Banerjee}
\author[1,2]{Subhasish Dutta Gupta}[orcid=0000-0002-7168-8549]
\ead{sdghyderabad@gmail.com}
\address[1]{Department of Physical Sciences, IISER-Kolkata, Mohanpur 741246, India}
\address[2]{Tata Institute of Fundamental Research, Hyderabad, Telangana 500046, India}

\begin{abstract}
We study theoretically near-symmetric resonant stratified media to show how a tiny broken spatial symmetry can effectively be probed by structured light with or without orbital angular momentum. This is achieved by examining both the in-plane and out of plane Goos-H\"anchen and Imbert Fedorov shifts, respectively, in the reflected light,  magnified by resonant enhancement and when needed with weak value amplification. We show that non-reciprocity in reflection for illumination from opposite ends can result in different shifts, even to the extent of shifts with opposite signs for tiny imbalance resulting from the broken symmetry. Our methodology is applicable to any general stratified medium with homogeneous and isotropic constituent layers being responsive to any entity that may cause such symmetry breaking - including temperature, refractive index, displacement etc. in the near symmetric case. This may have possible use for sensing applications.
\end{abstract}


\maketitle

\section{Introduction}

 The ubiquity of symmetry pervades all areas of physics with broken symmetry leading to novel discoveries \cite{Higgss1964,anderson1972,nambu2009,Rashba1}. There exists a great deal of work on optical systems probing the reciprocity in systems with broken symmetry \cite{potton2004,Dowling,labbook}. Note that reciprocity refers to identical outcome for reversal of light trajectory through the system. Problem gets much simplified with clear physical interpretations in case of a linear stratified medium. A very general consideration  \cite{Agarwal2002} revealed that transmission through such a system is always reciprocal independent of broken spatial and time-reversal (in the presence of loss or gain media) symmetry, while in nonlinear systems reciprocity may not hold leading to optical diode action \cite{Regularizationofthespectral,PhysRevB.97.205423}. In contrast, the reflection can exhibit a much richer variety even in a linear system \cite{Agarwal2002,Madhuri2012}. A system lacking spatial and time reversal symmetry can lead to nonreciprocity in reflection as was verified both experimentally and theoretically \cite{ArmitagePRB,Agarwal2002,VSCMangaRao2004,Armitage2014}. Note that nonreciprocity in reflection for an asymmetric lossy or lossless structure does not violate Lorentz reciprocity (see, for example, Section XII in \cite{Caloz2018PRA}). Transmission being always reciprocal, in full agreement with Lorentz reciprocity, can not lead to optical isolation or diode action, which essentially needs broken Lorentz reciprocity \cite{jalas2013NatPhotonics}. Even in a system without losses with broken spatial symmetry, the phase of the reflection coefficient can be nonreciprocal leading to sub and superluminal light for incident pulses \cite{MangaRao2004,labbook}. The nonreciprocity in phases for opposite propagation directions can have interesting consequences for incident beams, more so, for vector beams with or without orbital angular momentum. This is expected since for oblique incidence the beams can undergo in-plane or out of plane deflections (Goos-H\"anchen (GH) and Imbert Feodorov (IF) shifts, respectively \cite{Bliokh2013,artmann,golla2011goos}), which can be amplified by resonant modes of the stratified media \cite{Chen:07,Sensors2,Xu:24,homola2006surface}. In case of small shifts, they can further be enhanced by projective measurements invoking weak value amplification \cite{weak2,weak3,Mandira2014,XiongGHWVA2026,maiti2026}. 
\par
In this paper we present an in-depth theoretical study of propagation of vector beams through a stratified medium comprising of homogeneous and isotropic constituent layers with clear focus on the nonreciprocity of the beam shifts in reflection for opposite directions of illumination in an asymmetric structure. Hereafter, the word  \textit{nonreciprocity} will be used only in the context of the reflected wave, unless specified otherwise. We further explore its potential for sensing using the broken symmetry caused by tiny deviations (in displacement, RI mismatch, etc.) from an ideal symmetric structure. Of course the realization of an actual sensing device with ultra-high sensitivity will depend on many other important factors limiting the fabrication and detection schemes. The potential of nonreciprocity in GH shift in reflection to detect small displacements was first highlighted in an early paper \cite{Madhuri2012}. There have been other suggestions for GH-shift-based sensing mostly using surface plasmon polaritons, wave guide modes or Bloch surface waves \cite{homola2006surface,Xiaobo2006SPRsensor,Sensors2,Chen:07,Tang:25,TAYA2012204,Sensors1,KONG201862}. Along with the longitudinal GH shift, the transverse IF shift has emerged as an important platform for optical sensing and beam-shift engineering. In particular, electro-optically tunable IF shifts in graphene-based structures have demonstrated enhanced controllability through voltage-dependent optical responses \cite{ZHU2019165319}. Resonantly amplified IF shifts in plasmonic and double-prism configurations have also enabled highly sensitive optical detection and precision metrology applications \cite{Xu:24,Zhu:24}. Some very recent work focus on exploiting the nonreciprocity due to broken time reversal symmetry in reflection with Weyl semimetals \cite{weyl1,weyl2}.  However, most of this research is based on stationary phase approximation (Artmann formula \cite{artmann}), or at the best a scalar field description for the beams. Note that the stationary phase approximation breaks down at resonance due to sharp phase variation for high-Q modes. In contrast, here we present a full vectorial theory for the nonreciprocity in reflection resulting from the simplest form of broken spatial symmetry, enabling the extraction of both GH and IF shifts for beams without/with orbital angular momentum (OAM). We must add that our rigorous analysis is applicable to any asymmetric structure with no limitation on the nature (displacement, RI mismatch, etc) and magnitude of the asymmetry. In fact, it is applicable to any stratified medium, comprising of homogeneous and isotropic constituents.
\begin{figure}
	\centering
    \includegraphics[width=0.45\textwidth]{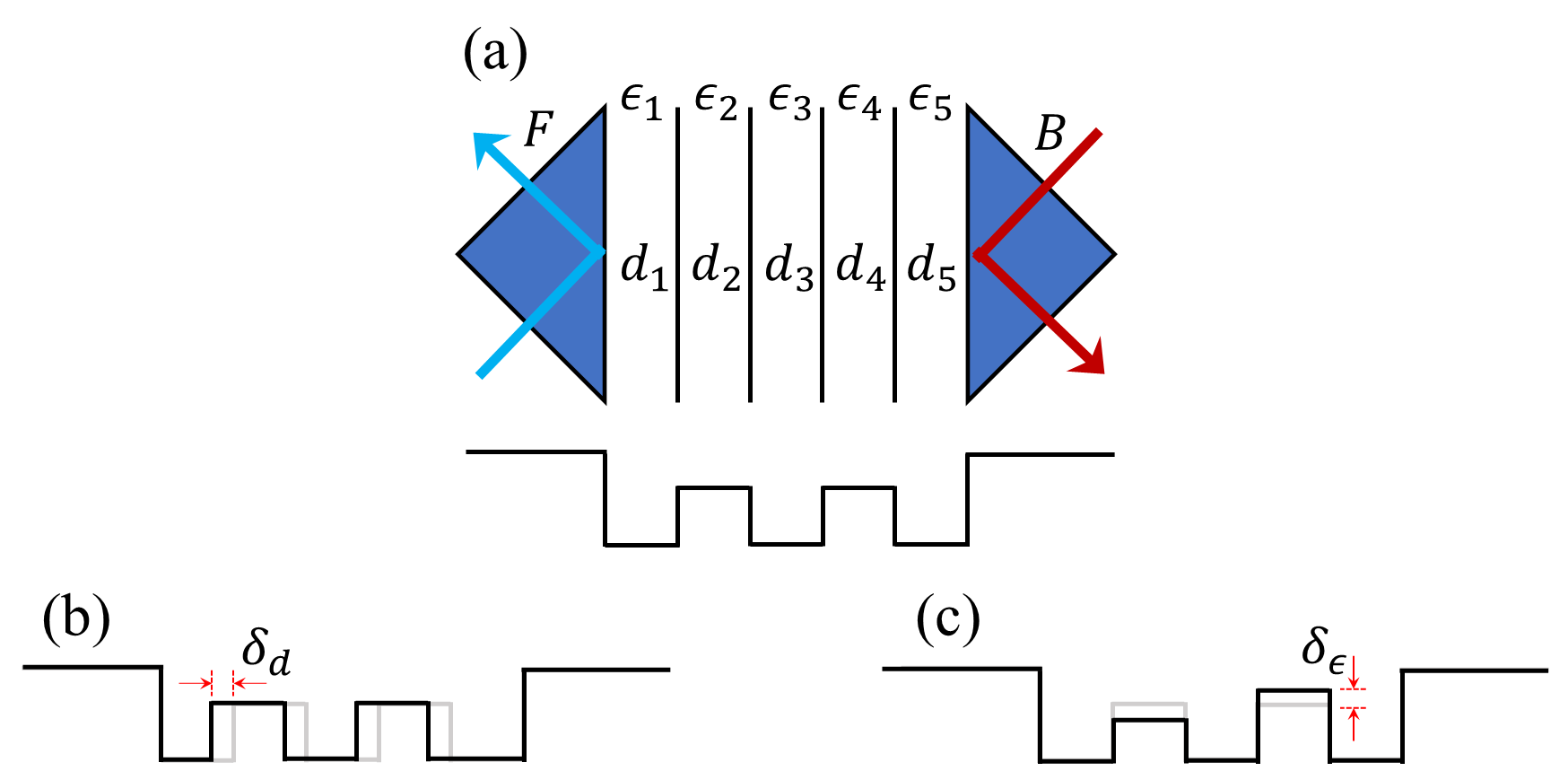}
	\caption{(a) Schematic view of the symmetric coupled waveguide structure with $\epsilon_i=\epsilon_f$, $\epsilon_1=\epsilon_3=\epsilon_5=\epsilon_s$, $\epsilon_2=\epsilon_4=\epsilon_g$, $d_1=d_5=d$, $d_2=d_4=d_g$. Layers with widths $d_2$ and $d_4$ are high index guiding layers separated by a low index spacer layer with width $d_3$. (b) Same as in Fig.~\ref{FIG:1}(a), except that now $d_1=d-\delta_d$, $d_5= d+\delta_d$ resulting in a translation of the coupled waveguides towards left. (c) Same as in Fig.~\ref{FIG:1}(a), except that now $\epsilon_2=\epsilon_g-\delta_{\epsilon}$, $\epsilon_4=\epsilon_g+\delta_{\epsilon}$ leading to refractive index mismatch between the guiding layers. The middle spacer layer mediates the coupling between the guides. Forward (backward) propagation is depicted by blue (red) arrows in Fig.~\ref{FIG:1}(a). $F~ (B)$ in the panel here and later refers to the forward (backward) incidence.}
	\label{FIG:1}
\end{figure}
\par
The structure of the paper is as follows. In Section \ref{sec2} we describe our system supporting coupled guided modes and the illumination geometry and outline the gist of angular spectrum method for beams. Sections \ref{sec3.1} and \ref{sec3.2} cover the results of numerical simulation for two cases, namely,  system with broken symmetry  caused by (a) translation and by (b) refractive index mismatch, respectively,
of the guiding layers. For reference we also include specimen results for the symmetric structure. Section \ref{sec3.3} presents the nonreciprocal effects in transverse IF and vortex-induced shifts for linearly polarized and OAM carrying beams under projective measurements to bring out the scope of weak value amplification. Finally, we conclude by summarizing our main findings.

\section{Theoretical Framework}\label{sec2}
Consider the folded Kretschmann-like configuration supporting coupled guided modes (see Fig.~\ref{FIG:1}(a)), which can be illuminated by light beams with beam waist $w_0$ and wavelength $\lambda$ incident at an angle $\theta$ from either end. Depending on the widths of the guiding layers the system, in principle, can support higher order modes, though we focus only on the fundamental mode. Further, the coupling induces mode splitting leading to long- and short- range modes \cite{longshotmode3,longshotmode1,longshotmode2,labbook}, out of which we pick the one with the narrowest resonance, which can offer the largest Q-factor with large local field enhancement. The structure in Fig.~\ref{FIG:1} (a) can be chosen to be symmetric or the symmetry can be broken by introducing a small (a) translation or (b) refractive index mismatch of the guiding layers as shown in Figs \ref{FIG:1}(b) and \ref{FIG:1}(c), respectively. These perturbations are designed to mimic realistic scenarios encountered in ultra-sensitive sensing applications \cite{KONG201862,Sensors1,Sensors2}, where minute structural or material variations can significantly influence the optical response. The introduction of such asymmetry breaks the spatial symmetry of the structure, leading to an asymmetric phase winding in reflection for opposite propagation directions, which gives rise to nonreciprocal beam shifts.
\par

In order to analyze the beam shifts, we employ the generalized angular-spectrum framework introduced by Bliokh and Aiello~\cite{Bliokh2013}. In this approach, an optical beam is represented as a coherent ensemble of plane-wave components whose wave vectors are distributed around the central propagation direction. Our group has extended this formalism to accommodate a wider transverse momentum distribution \cite{maiti2026,SinhaBiswas2023,BISWAS2024130766}, allowing the treatment of beams beyond the conventional paraxial regime and providing a generalized description of their propagation through stratified media. Within this framework, the incident beam profile is first decomposed into its Fourier spectrum. For a Gaussian beam propagating along the $z$-axis of the beam coordinate system, is written as~\cite{Bliokh2013}
\begin{equation}
\left|\mathbf{E}_i\right\rangle
=
\frac{w_0}{\sqrt{2 \pi}}
\exp\left[-\left(k_x^2+k_y^2\right)w_0^2/4\right]
\left(A_p\mathbf{e}_p+A_s\mathbf{e}_s\right),
\label{eq2}
\end{equation}
 while $A_p$ and $A_s$ specify the amplitudes of the orthogonal polarization components. For each off-axis spatial-frequency component, the corresponding polar and azimuthal angles are determined from~\cite{PB_phase_PRL}
\begin{equation}
\theta_i =
\tan^{-1}\left[
\frac{
\sqrt{k_y^2+\left(k_x\cos\vartheta_i+k_z\sin\vartheta_i\right)^2}
}{
-k_x\sin\vartheta_i+k_z\cos\vartheta_i
}
\right],
\label{eq3}
\end{equation}
\begin{equation}
\phi_i =
\tan^{-1}\left(
\frac{k_y}
{k_x\cos\vartheta_i+k_z\sin\vartheta_i}
\right),
\label{eq4}
\end{equation}
where
\begin{equation}
k_z=\sqrt{k_0^2-k_x^2-k_y^2}.
\end{equation}

We further consider Laguerre Gaussian (LG) beams carrying orbital angular momentum, given by~\cite{Bliokh2013}
\begin{equation}
\label{eq5}
\begin{split}
\left|\mathbf{E}_i\right\rangle
\propto {}&
\frac{w_0}{\sqrt{2\pi}}
\exp\left[-\left(k_0w_0\right)^2\theta_z^2/4\right]
\theta_z^{|\ell|}
e^{i\ell\phi+i k_0\left(1-\theta_z^2/2\right)z}
\\
&\times
\left(A_p\mathbf{e}_p+A_s\mathbf{e}_s\right),
\end{split}
\end{equation}
where
\begin{equation}
\theta_z=\frac{\sqrt{k_x^2+k_y^2}}{k_0},
\qquad
\phi=\tan^{-1}\left(\frac{k_y}{k_x}\right),
\end{equation}
and $\ell=0,\pm1,\pm2,\ldots$ denotes the topological charge of the vortex.
For each spatial harmonic, the corresponding Fresnel coefficients and coordinate transformations are applied to obtain the reflected angular spectrum. The reflected field in the momentum-space representation can therefore be expressed as~\cite{Bliokh2013,SinhaBiswas2023}
\begin{equation}
|\mathbf{E}_r\rangle
=
U_r^{\dagger}F_rU_i|\mathbf{E}_i\rangle,
\label{eq6}
\end{equation}
where
\begin{equation}
U_r=
\hat{R}_y(\theta_r)
\hat{R}_z(\phi_r)
\hat{R}_y(-\vartheta_r),
\label{eq7}
\end{equation}
and
\begin{equation}
F_r=
\begin{pmatrix}
r_p & 0\\
0 & r_s
\end{pmatrix}.
\label{eq8}
\end{equation}
Here, the subscripts $i$ and $r$ denote the incident and reflected fields, respectively, while $\hat{R}_y$ and $\hat{R}_z$ represent rotations about the $y$- and $z$-axes in the laboratory frame. Finally, the spatial distribution of the reflected beam is obtained by performing the inverse Fourier transformation of Eq.~(\ref{eq6}). The GH and IF shifts are then evaluated from the corresponding reflected intensity distribution by calculating the displacement of its intensity centroid. This procedure enables the beam shifts to be obtained directly from the full reflected beam profile rather than from a first-order phase expansion alone. 

\begin{figure}
	\centering
	\includegraphics[width=.45\textwidth]{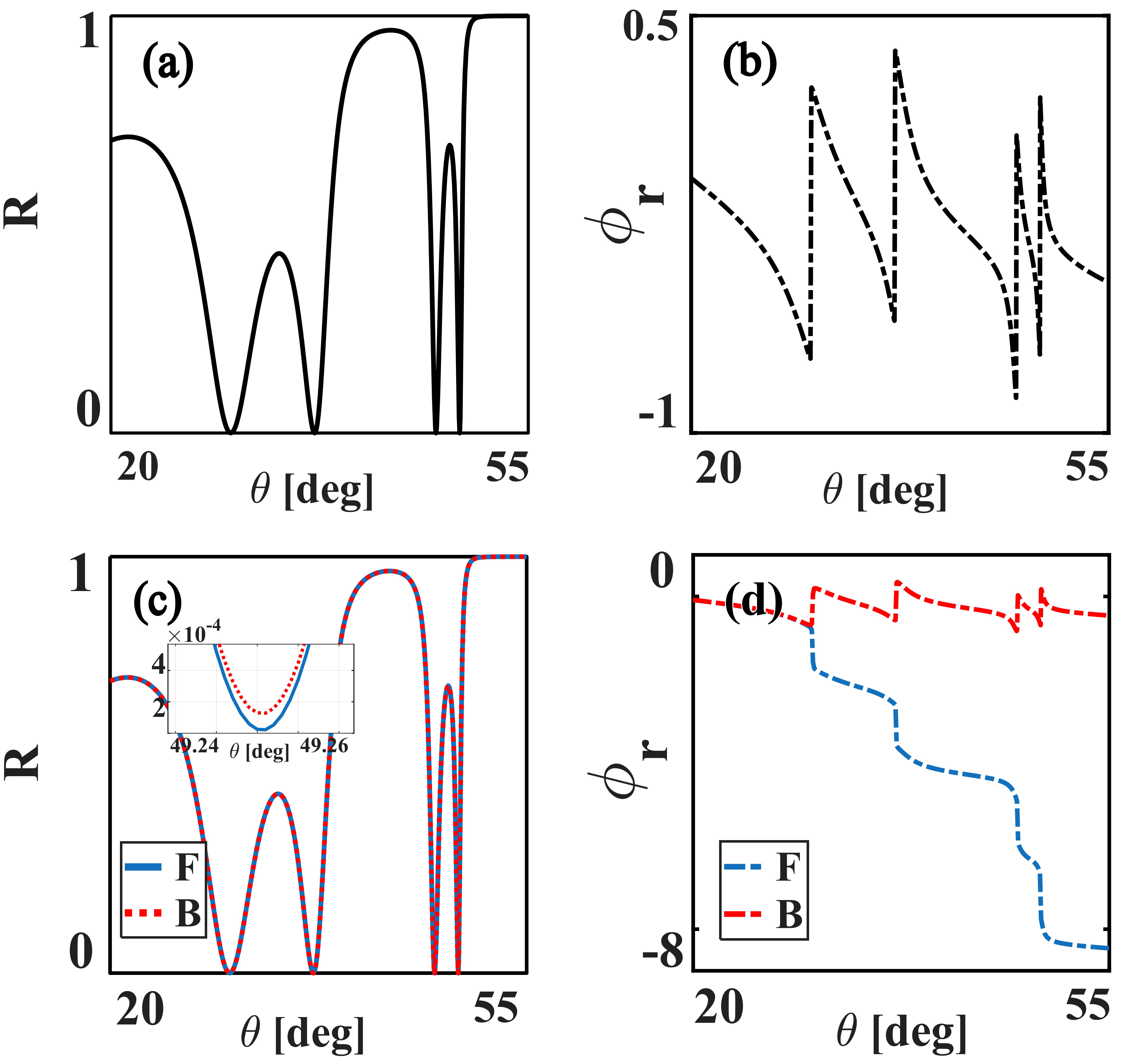}
	\caption{Plane wave results for s-polarized light: (a) Intensity reflection coefficient $R =|r|^2$ and (b) phase $\phi_r$ (in units of $\pi$) of $r$ as functions of $\theta$ for the symmetric structure with $d=0.1\mu m$. Lower panels (c) and (d) show the same except that now $d_1=d-\delta_d$ and $d_5=d+\delta_d$ with $\delta_d=0.0005\mu m$. All the panels show the results for both forward and backward propagation. The inset in (c) shows an expanded portion highlighting the non-reciprocity in intensity reflection due to broken space and time reversal symmetry. Other parameters are as follows $d_g=0.3201\mu m$, $d_3=0.095\mu m$, $\epsilon_g=3.9085+i* 0.0001$ ,$\epsilon_s=2.2883$, $\lambda=0.69~\mu m$. Forward (backward) propagation is depicted by blue (red) curves.}
	\label{FIG:2}
\end{figure}
\begin{figure}
	\centering
	\includegraphics[width=.48\textwidth]{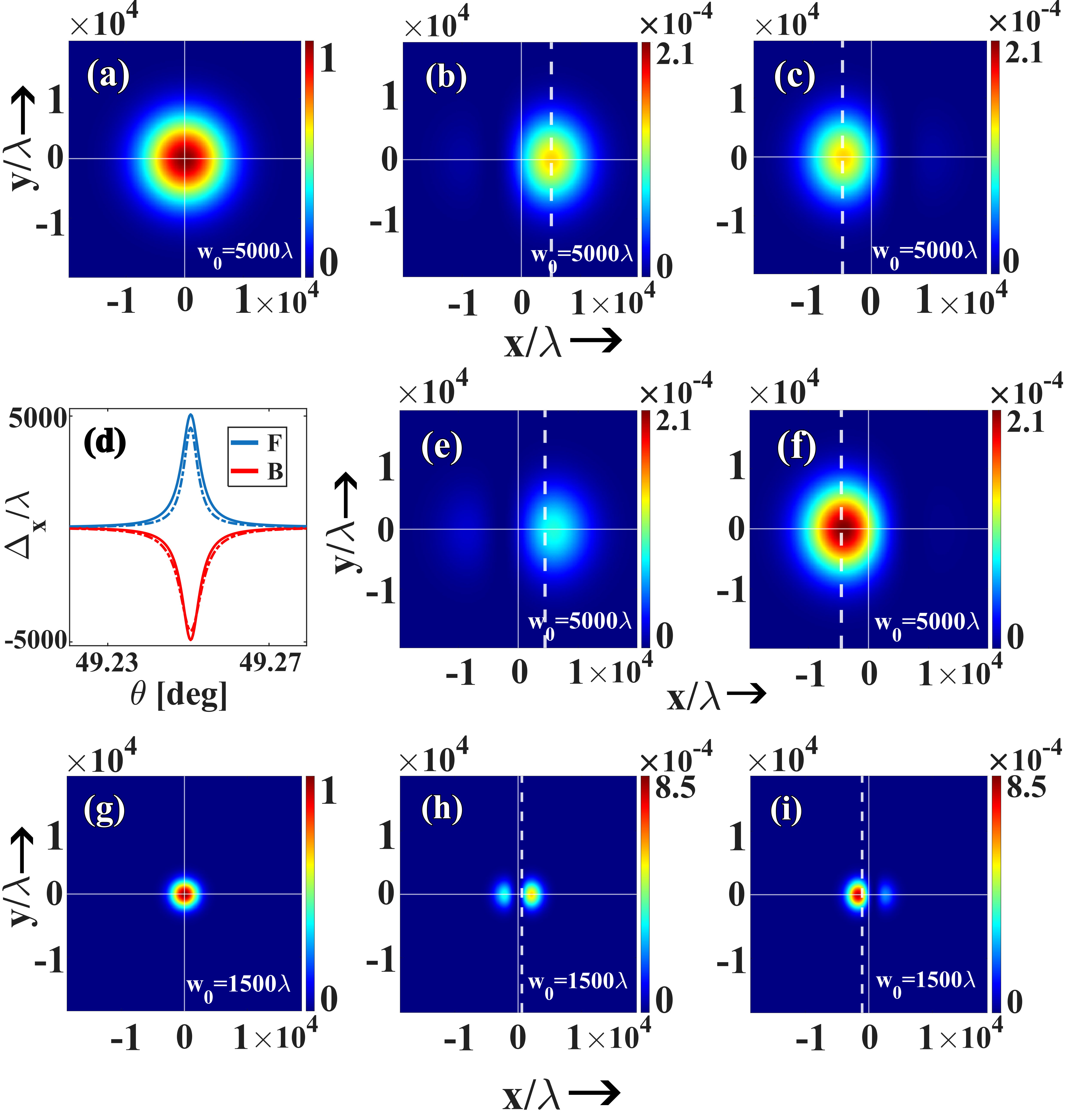}
	\caption{Incident \textit{s}-polarized Gaussian beams with beam waists (a) $w_0=5000 \lambda$ and (g) $w_0=1500 \lambda$ illuminating the asymmetric structure shown in Fig.~\ref{FIG:1}(b) at an angle $\theta=49.2506 ^\circ$. (b),(e),(h)  ((c),(f),(i)) show the reflected beam profiles for forward (backward) illumination. (b),(c) ((e), (f)) are for lossless $\Im(\epsilon_g)=0$ (lossy ($\Im(\epsilon_g)=0.0001)$)) structure with waist $w_0=5000\lambda$. (d) shows GH shift as function of $\theta$ where solid (dashed) lines are for lossless (lossy) structure. (h),(i) are for the structure with loss with $w_0=1500\lambda$. The solid and the dashed white lines mark the centroids of the incident and reflected beams, respectively. Other parameters are as in Fig. \ref{FIG:2}.}
	\label{FIG:3}
\end{figure}
\section{Results}\label{sec3}
We first analyze the optical response of the symmetric resonant tunneling structure depicted in Fig. \ref{FIG:1}(a) for $s$-polarized plane wave excitation. The geometrical and material parameters are chosen as $d=0.1 ~\mu m$, $d_3=0.095 ~\mu m$ and $d_g=0.3201~\mu m$, with corresponding permittivities $\epsilon_s=2.2883$ and $\epsilon_g=3.9085$. All the media are assumed to be nonmagnetic. Excitation of the coupled modes is mediated by identical high-index prisms characterized by $\epsilon_i=\epsilon_f=6.145$. We have included losses in the guiding layers by a finite but small imaginary part of $\epsilon_g$ to read out the signatures of nonreciprocity when both spatial and time-reversal symmetries are broken. The results for the intensity reflection coefficient $R$ and phase $\phi_r$ are shown in Figs \ref{FIG:2}(a) and \ref{FIG:2}(b), respectively, for the fundamental and the first order modes. The well-understood splitting for all the orders can be seen in these figures leading to the narrowest resonance at $\theta=49.251$ with the corresponding sharpest phase jump. Note that such high-$Q$ modes are essential to enhance the sensitivity of a generic sensor. As expected, because of the inherent symmetry, reciprocity both in amplitude and phase is not violated. We now show how the broken symmetry can drastically affect the outcome for opposite directions of propagation not only for the reflection amplitudes and phases but also for the the in-plane and out of plane beam shifts. The non-reciprocity persists not only for plane waves but also for general vector beams without or with OAM. We start with a brief but general analysis of phases in asymmetric structure for opposite directions of illumination. Details of the phase behavior for an asymmetric stratified medium was studied in the context of Wigner delay for optical pulses
(see Eq.(6) and the discussions thereafter in \cite{MangaRao2004}) It was shown that sign-wise these phases are complementary, which may result in opposite signs of the Wigner phase time. In the framework of a stationary phase approximation, similar conclusions hold for beam shifts as well, since frequency derivative of phase (Wigner delay) is replaced by the negative of the derivative of phase with respect to the surface component of the wave vector (GH shift). Thus, translated to the space domain, the essense of Eq.(6) in \cite{MangaRao2004} reduces to opposite signs of GH shift ($ \sgn(x_{GH}^F)=-\sgn(x_{GH}^B)$). Hereafter, ample examples illustrating the direction dependence of phases and the shifts are given in the following.
 
\begin{figure}
	\centering
	\includegraphics[width=.5\textwidth]{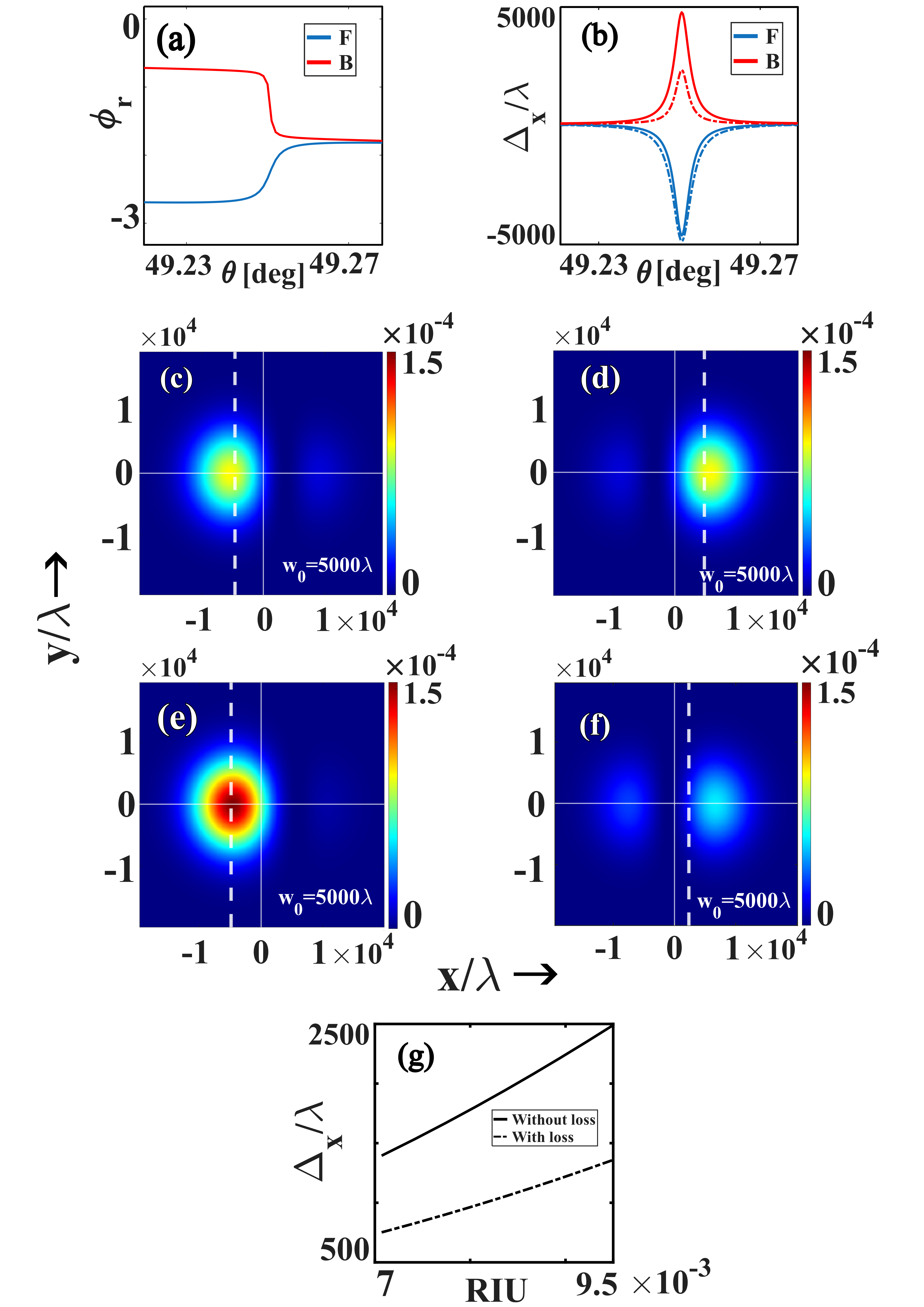}
	\caption{Results for RI mismatch induced asymmetry: (a) Phase $\phi_r$ (in units of $\pi$) of $r$ as functions of $\theta$ for $\epsilon_2=\epsilon_g-\delta_{\epsilon}$ and $\epsilon_4=\epsilon_g+\delta_{\epsilon}$ with $\delta_\epsilon=0.0005$. (b) GH shift as a function $\theta$ for both forward and backward propagation for lossless(solid line) and lossy(dashed line) systems. (c) and (d) ((e), (f)) represent beam profiles for the forward and backward directions for lossless (lossy) systems. (g) Differential GH shift $\Delta_x$ between GH shifts for forward and backward illumination as a function of RI change for a structure without (solid) and with loss (dashed). Corresponding linear fits  reveal the sensitivity as $3.2066 \times 10^{-9}$ RIU/nm and $5.7799 \times 10^{-9}$ RIU/nm for lossless and lossy structures, respectively. Other parameters are as in Fig.\ref{FIG:2}.}
	\label{FIG:4}
\end{figure}

\subsection{Symmetry breaking with translation}\label{sec3.1}
Now, we draw attention to the asymmetric structure shown in Fig. \ref{FIG:1}(b), where the spatial symmetry is broken by a small translation $\delta_d=0.0005~\mu m$ of the coupled guides keeping the total optical width (as in the symmetric case) and other parameters fixed. Again, for reference, the results for the plane wave reflectivity and phase are shown in Figs \ref{FIG:2}(c) and \ref{FIG:2}(d), respectively. As can be seen from these figures, the response is now nonreciprocal mostly due to the contrasting behavior of phases in a lossy/lossless system for forward and backward propagation. We assumed the operating wavelength to be far away from any of the resonances of the dielectric, justifying the presence of a phenomenological tiny loss given by $\Im(\epsilon_g)=0.0001$. Monochromatic character of the beam did not necessitate the use of any specific model for dispersion/loss since the target was to explore the broken reciprocity of reflected intensity $R$. Even for small displacement, tiny losses lead to nonreciprocity in $R$, while it would be identical in absence of losses ($\Im(\epsilon_g)=0$) (see inset in Fig.~ \ref{FIG:2}(c)). Note that phase response (Fig.~ \ref{FIG:2}(d)) remains nearly the same for lossy and lossless (not shown) systems with broken symmetry. Going by the Artmann formula or more accurate analytical expressions of Bliokh and Aiello \cite {artmann, Bliokh2013} the opposite slopes of the phase curves predict distinct signs for the GH shifts for forward and backward propagation. These preliminary plane wave results have profound implications for beams since a beam can be considered as a collection of plane waves. It is also important to consider the effects of the beam spot size which has a critical effect on the reflected beam profiles.
In what follows, we present the results for reflected beam shapes using the angular spectrum decomposition.
\par


\begin{table*}[h]
    \centering
    \begin{tabular}{l l l l}
        \hline
        Type & Sensitivity reported in paper & Converted in `RIU/nm' unit & Paper \\
        \hline
        Theoretical GH/BIC & $3.58 \times 10^{6}\, \mu\text{m}/\text{RIU}$ & $2.79 \times 10^{-10}~\text{RIU}/\text{nm}$ & \cite{Jiang2023Micromachines} \\
        Theoretical PCF-SPR & $1.2 \times 10^{5}\, \text{nm}/\text{RIU}$ & $8.13 \times 10^{-6}~\text{RIU}/\text{nm}$ & \cite{Al2024IEEEAccess} \\
        Experimental GH/BSW & $2.0 \times 10^{-9}~\text{RIU}/\text{nm}$ & $2.0 \times 10^{-9}~\text{RIU}/\text{nm}$ & \cite{kong2018SensorsActuators} \\
        Experimental Au-SPR & $7166.27\, \text{nm}/\text{RIU}$ & $1.395 \times 10^{-4}~\text{RIU}/\text{nm}$ & \cite{fu2025infraredphysicstechnology} \\
        Experimental Au-SPR & $1.82 \times 10^{-8}\,\text{RIU}/nm$ & $1.82 \times 10^{-8}~\text{RIU}/\text{nm}$ & \cite{Xiaobo2006SPRsensor} \\
        Experimental RR & $112.5 \ nm/\text{RIU}$ & $8.89 \times 10^{-3}~\text{RIU}/\text{nm}$ & \cite{butt2024AppliedSciences} \\
        Our prediction & $5.7799\times 10^{-9} ~\text{RIU}/nm$ & $5.7799 \times 10^{-9}~\text{RIU}/\text{nm}$ &  \\
        \hline
    \end{tabular}
    \caption{Recent developments in the RI sensing platforms.GH: Goos-H\"anchen, BIC: Bound states in continuum, PCF: Photonic Crystal Fiber, SPR: Surface plasmon resonance, BSW: Bloch surface wave, RR: ring resonator.}
    \label{tab:placeholder_label}
\end{table*}

Reflected beam shapes for the forward and backward illumination are shown in Fig.~\ref{FIG:3} for two different values of the beam waist, namely, $w_0=5000 \lambda$ and $1500 \lambda$ (compare third row Fig.~\ref{FIG:3} with first and second). In order to bring out the effect of losses, we have presented the results for lossless coupled guides in the figs \ref{FIG:3}(b) and \ref{FIG:3}(c). The common feature that emerges from a comparison of forward and backward propagation is the opposite signs of the GH shift, while the shift is more pronounced for larger spot size. However, resonant enhancement facilitates giant shifts comparable to its spot size. Recall that a larger spot size corresponds to loose focusing approaching the plane wave results while tighter focusing may lead to distortion of the beam with weaker additional lobe (clearly seen in Figs. (e),(f), (h),(i)) in the presence of losses with direction-dependent strength. For example the side lobe is extremely weak for backward propagation in Figs.~ \ref{FIG:3}(f) and \ref{FIG:3}(i) in contrast to the excitation from the other side. Losses play a vital role due to the degradation of the quality factor leading lower values of the shifts which one can observe in Fig.~\ref{FIG:3}(d) (also see Figs.~ \ref{FIG:3}(b), (c) and Figs.~ \ref{FIG:3}(e), (f)). It is pertinent to note that, counterintuitive to general perception, losses can lead to more intense reflected beam for backward propagation, as compared to the reflected beam for the symmetric case  (compare Figs.~ \ref{FIG:3}(c) and \ref{FIG:3}(f)), while it leads to suppression for forward propagation. Thus losses can imbalance the beam intensity as compared to the equal distributions for lossless systems. This is a direct consequence of the nonreciprocity in reflected intensity in the lossy asymmetric structure.

\subsection{Symmetry Breaking with refractive index mismatch}\label{sec3.2}
We now turn to the system shown in Fig.~\ref{FIG:1}(c), where a small index mismatch $2\delta_{\epsilon}$ has been introduced to break the spatial symmetry keeping the total optical width the same as in the symmetric case. For the system in Fig.~\ref{FIG:1}(c) the plane wave results for phases for the forward and backward directions are shown in Fig \ref{FIG:4}(a). Opposite in sign GH shifts can be inferred from Fig \ref{FIG:4}(a). The Rest of the panels in Fig \ref{FIG:4} present results using the angular spectrum method. Figs. \ref{FIG:4}(b) compares results for lossless and lossy guides demonstrating the effects of losses on the GH shifts. Indeed the GH shifts are modified considerably in a direction-dependent manner in presence of losses even for a tiny RI mismath of $\delta_{\epsilon}=0.0005$. Interestingly, positive (negative) shifts gets subdued (enhanced) in presence of losses. This is in contrast to the results in Fig.~\ref{FIG:3}(d), where for both the directions GH shift is reduced. Next four panels present the beam shapes for forward and backward directions as well as compare the shifts for lossless and lossy systems. Direction-dependent features like in Figs. \ref{FIG:3} (b), (c), (e) and (f) are observed, albeit now the  beam is more intense for forward direction as compared to the symmetric case. 
\par
We now explore the possible use of our results for sensing applications. Note that presented results were obtained for tiny deviations from an idealized perfect symmetric system. The universally accepted figure of metrit for a RI sensing device is its sensitivity expressed in units of RIU/nm. We present a theoretical estimate for the RI mismatched case by looking at the differential GH shift (difference in the GH shifts for forward and backward propagations) $\Delta_x=x_{GH}^F-x_{GH}^B$ as a function of refractive index unit. Fig. \ref{FIG:4}.(g) shows the same for lossless and lossy structures. The corresponding sensitivity in units of RIU/nm is estimated from an inverse slope of the linear fit of these plots like in \cite{KONG201862}. Loss-induced degradation of the sensitivity can easily be read from this plot. Nevertheless, it is clear that one can extract very high theoretical sensitivity $\sim 10^{-(8-9)}$ RIU/nm from such simple systems exploiting the nonreciprocity caused by the asymmetry induced by the material to be probed. It must be stressed that the practical realization of such a sensor is highly demanding and beyond the scope of our paper, since one has to take into account fabrication and measurement limitations, alignment, dispersion and bandwdith constraints, not to mention linewidth fluctuations of the source and actual signal to noise features of the detection system. Nevertheless, the theoretical estimate for the idealized case is lucrative enough to initiate further research exploiting our ideas. Table 1 (though not exhaustive) summarises some old and recent developments in sensors making this point very clear. As can be seen from Table 1, our theoretical estimate compares well with other sensing schemes.

\subsection{Nonreciprocity in IF and vortex-induced shifts}\label{sec3.3}
We now focus on the influence of symmetry breaking on the IF shift under different polarization measurement schemes. For conventional IF shift measurements using incident left- or right-circularly polarized Gaussian beams (results not shown), the difference between the forward and backward propagating beams remains relatively small and is often difficult to resolve experimentally. This motivates the exploration of alternative polarization-selective strategies capable of amplifying the underlying nonreciprocal transverse displacement.
\begin{figure}
	\centering
	\includegraphics[width=0.45\textwidth]{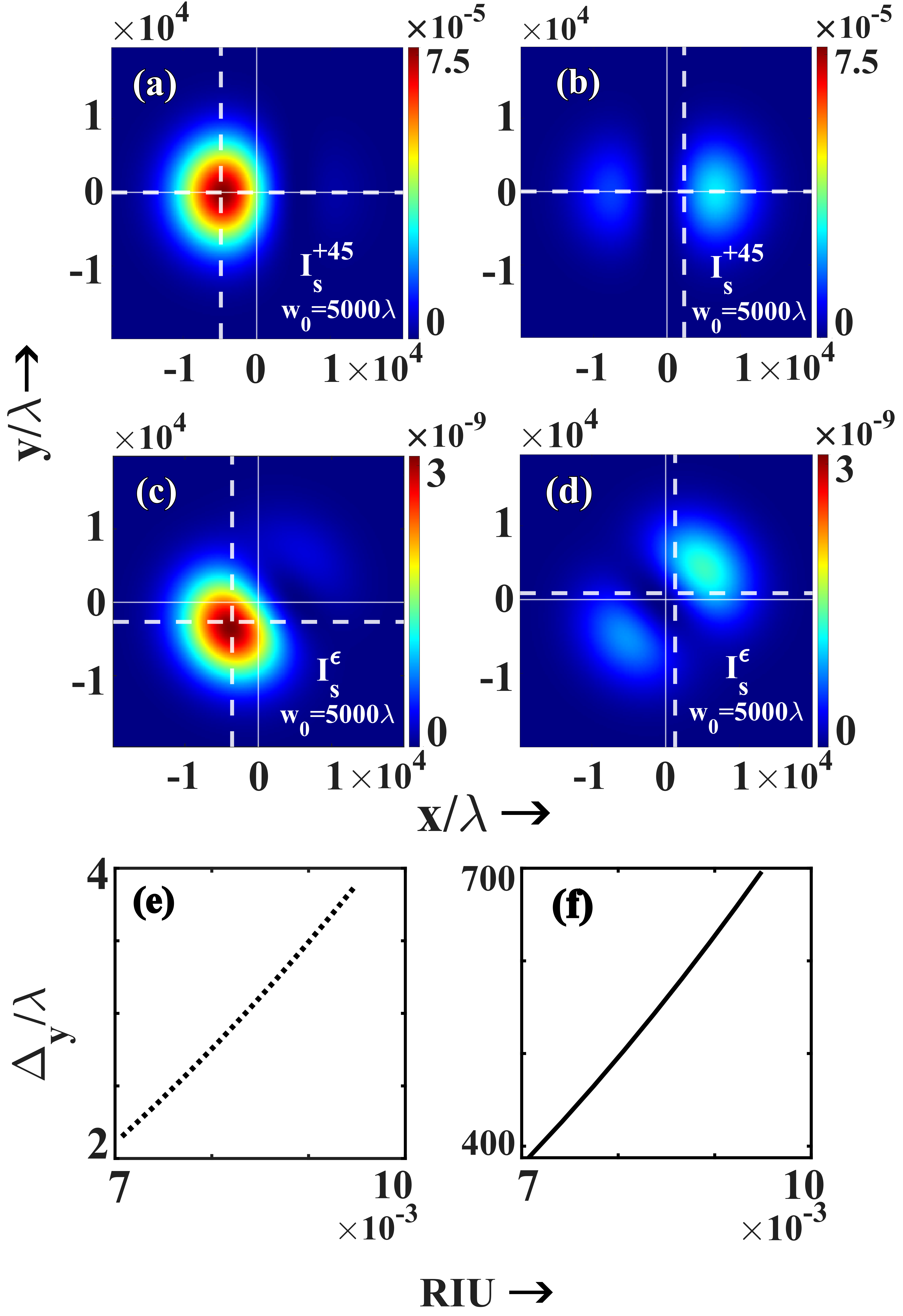}
	\caption{Results on non-reciprocal IF shift Induced by RI mismatch: (a) and (b) show the reflected beam profiles for forward and backward illumination, respectively, for an incident \textit{s}-polarized beam projected onto the $+45^\circ$ polarization state. (c) and (d) present the corresponding beam profiles when the reflected beam is projected onto a near-orthogonal polarization state with offset parameter $\epsilon=0.004$. (e) and (f) depict the differential transverse shift $\Delta_y$, defined as the difference between the IF shifts for forward and backward propagation, as a function of RI variation. (e) corresponds to the $+45^\circ$ projection scheme, while (f) represents the near-orthogonal projection configuration. Linear fit of the numerical data reveals sensitivities of $2.0183 \times 10^{-6}$ RIU/nm and $1.1304 \times 10^{-8}$ RIU/nm for $+45^\circ$ projection and near-orthogonal weak-value amplification scheme, respectively. Other parameters are as in Fig.\ref{FIG:2}.}
	\label{FIG:5}
\end{figure}
\begin{figure}
	\centering
	\includegraphics[width=0.45\textwidth]{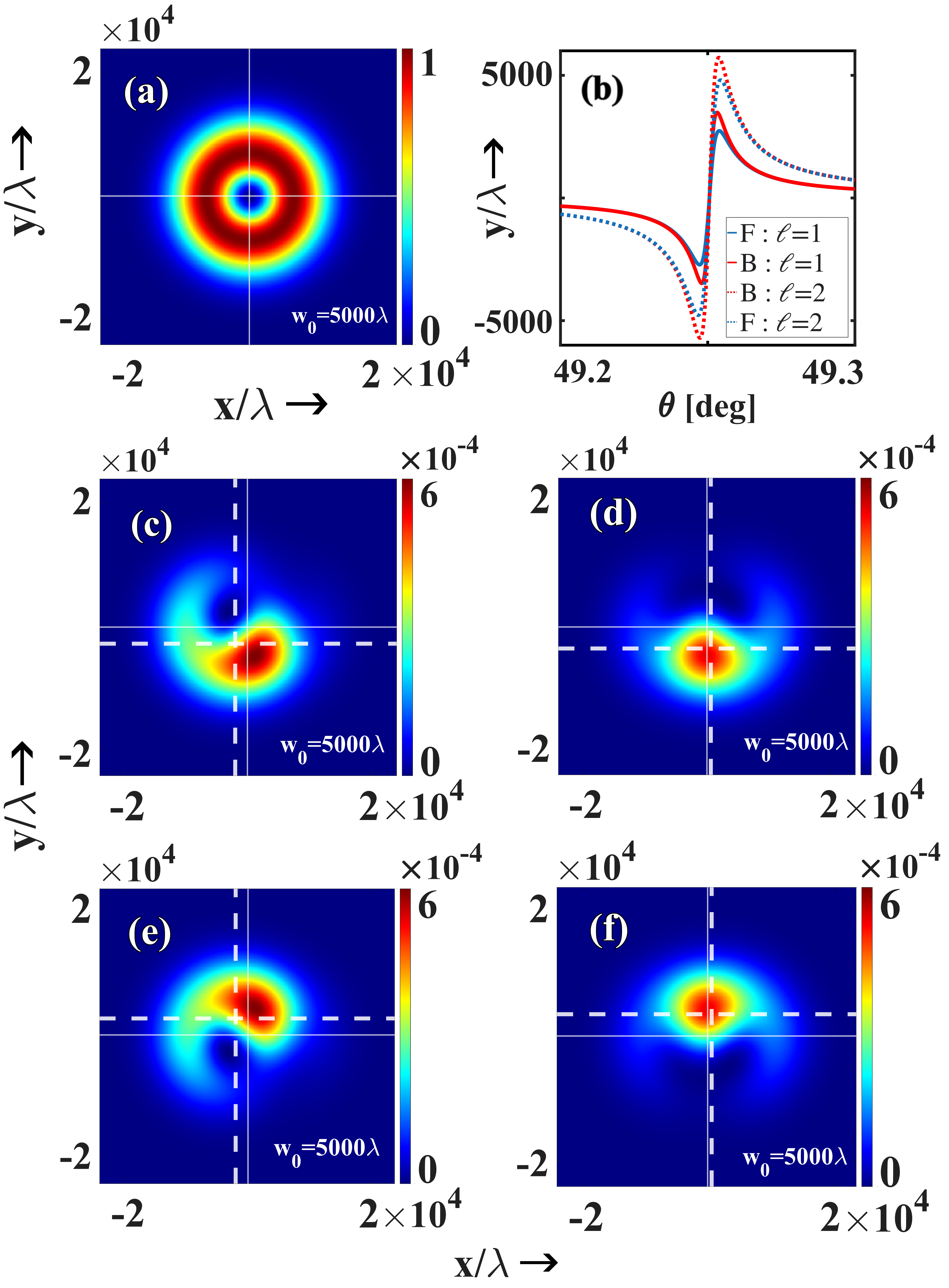}
	\caption{Results for \textit{s}-polarized LG beam: Panel (a) shows the intensity distribution of the incident LG beam. (b) Presents the IF shift as a function of the angle of incidence for forward (blue) and backward (red) propagating beams. Panels (c) and (d) depict the reflected intensity profiles for forward and backward propagation, respectively, when the angle of incidence is tuned at the dip of the  plot(b) $\theta=49.2478^\circ$. Panels (e) and (f) show the corresponding reflected beam profiles when the incidence angle is tuned at the peak of the plot(b) $\theta=49.2533^\circ$. All reflected LG beam profiles shown here are calculated for a topological charge of $\ell=1$. Other parameters are as in Fig.\ref{FIG:2}.}
	\label{FIG:6}
\end{figure}
\subsubsection*{Projective Measurements}
Recall that finite-width beam consists of a distribution of wave vectors surrounding the central wave vector. Upon reflection, each component experiences a different polarization transformation, resulting in a reflected beam that is no longer uniformly polarized. This spatially varying polarization distribution gives rise to markedly different centroid shifts when different polarization components of the reflected field are selectively analyzed \cite{weak3,weak4}. We first consider a projective measurement scheme in which an incident \textit{s}-polarized Gaussian beam is projected onto the $+45^\circ$ polarization state after reflection. The strong resonant suppression of the reflection coefficient $r$ gives rise to a small transverse IF shift. The corresponding reflected beam profiles for forward and backward illumination are displayed in Figs.~\ref{FIG:5}(a) and \ref{FIG:5}(b), respectively.
Note that one has a more intense reflected beam for the forward propagation as compared to that for the backward case. In order to amplify the nonreciprocal response, we employ a weak-value amplification \cite{weak1,weak2,weak3,weak4} protocol. The incident beam is prepared in the $s$-polarized state $(0,1)^{T}$ and the reflected beam is post-selected onto a nearly orthogonal state $(1,\epsilon)^{T}$. In this regime, the weak spin orbit interaction between the polarization and spatial degrees of freedom is dramatically amplified through near-orthogonal post selection \cite{maiti2026}, resulting in a giant transverse displacement depicted in Figs.~\ref{FIG:5}(c) and \ref{FIG:5}(d). To quantify the sensing performance, we calculate the differential transverse shift
$\Delta_y=y^F_{IF}-y^B_{IF}$. Similar to the methodology adopted in Sec. 3.2, the sensitivity is extracted from the slope of the differential shift as a function of refractive-index variation. The resulting differential shifts for the $+45^\circ$
 projection scheme and the weak-value amplification protocol are shown in Figs. \ref{FIG:5}(e) and \ref{FIG:5}(f), respectively. Clearly, a notable enhancement is observed when weak-value amplification is employed ($1.1304 \times 10^{-8}$ RIU/nm). These theoretical results demonstrate that even a relatively simple symmetry-broken resonant structure, when combined with polarization-selective amplification techniques, may offer sensitivity. comparable with recent IF-shift-based sensing platforms \cite{ZHU2019165319,Xu:24}. For brevity, we have restricted the present discussion to RI asymmetry. However, an analogous nonreciprocal IF response can also be obtained through translation-induced symmetry breaking, exhibiting qualitatively similar behavior and sensing characteristics.

 \subsubsection*{Vortex Induced Shifts}

We further investigate the response of OAM carrying Laguerre Gaussian (LG) beams in the presence of a small RI mismatch, following the procedure described in Sec.~\ref{sec3.2}. The IF shifts for forward and backward propagation are calculated for LG beams with topological charges $\ell=1$ and $\ell=2$, as shown in Fig.~\ref{FIG:6}(b). Unlike the Gaussian beam case, the LG beams do not exhibit opposite-sign IF shifts at a common incidence angle. The observed dispersion-like behavior becomes more pronounced with increasing $\ell$, reflecting the coupling between spatial and angular GH/IF shifts induced by the optical vortex \cite{LG2,LG1}. Consequently, LG beams are less suitable for the differential sensing scheme based on measurements at a common angle. For completeness, we also examine the reflected beam profiles at the dip and peak positions of the shift curves. As shown in Fig.~\ref{FIG:6}(c)--(f), the incident doughnut-shaped profile undergoes substantial distortion near resonance. This arises from the strong spectral filtering of the resonant multilayer, which modifies the amplitude and phase of different spatial harmonics and consequently deforms the vortex structure \cite{maiti2026}. These observations further highlight the complex interplay between resonance, OAM, and symmetry breaking in determining the beam shift characteristics of structured light.

\section{Conclusions}\label{con}
In conclusion, we have presented a theoretical study of the nonreciprocal effects in both GH and IF shifts of the reflected beam when the spatial and temporal symmetry of a coupled waveguide structure is broken. The analysis,  based on a rigorous, fully vectorial angular spectrum method for beams with or without angular momentum is applicable to any stratified medium with homogeneous and isotropic constituents. The full vectorial approach enabled us to obtain the reflected beam profiles in addition to the beam-shift values, allowing finite beam waist effects to be examined directly. To be specific, tighter focusing was shown to introduce beam distortion and emergence of side lobes. Nonreciprocity in a lossy asymmetric system was shown to lead to an interesting counterintuitive direction dependent enhancement/inhibition of the reflected beam intensity compared to that for the symmetric case. Most importantly, for tiny deviations (caused by translation or RI mismatch) from an ideal symmetric structure, was shown to lead to giant differential GH shift for opposite directions of propagation of linearly polarized Gaussian beams. We also reported how a small IF or vortex induced shift can be magnified by means of projective measurements exploiting weak value amplification for near orthogonal projection. We explored further the possible use of our scheme for sensing the tiny RI mismatch causing the asymmetry in the ideal symmetric system, reporting a theoretical estimate of $~10^{-9}$RIU/nm, which compares well with other schemes. As the present work is primarily theoretical for an idealized symmetric structure, we have not explicitly incorporated the various sources of uncertainty including fabrication and detection limitations, that would inevitably arise in a practical sensing device. Nevertheless, we hope that our theoretical estimate of lucrative high sensitivity will motivate other groups to take up further research based on our suggestion.



%

\end{document}